\documentclass[article]{IEEEtran}
\IEEEoverridecommandlockouts
\usepackage{multirow,multicol}
\usepackage{graphicx}
\usepackage{graphics}
\usepackage{amsfonts}
\usepackage{algorithm}
\usepackage{amssymb}
\usepackage{amsmath}
\usepackage{url}
\usepackage{lipsum}
\usepackage{mathrsfs}
\usepackage{subfigure}
\usepackage{savesym}
\savesymbol{iint}
\restoresymbol{TXF}{iint}
\usepackage{bm}
\usepackage[noend]{algpseudocode}

\makeatletter
\def\BState{\State\hskip-\ALG@thistlm}
\makeatother

\makeatletter
\newlength \figwidth
\makeatother

\begin{document}

\title{DRAG: Deep Reinforcement Learning Based Base Station Activation in Heterogeneous Networks\vspace{-0.0cm}}
\author{
\IEEEauthorblockN{
Junhong YE,~\IEEEmembership{Student~Member,~IEEE}, Ying-Jun~Angela~Zhang,~\IEEEmembership{Senior~Member,~IEEE}
}
}

\maketitle

\vspace{-0.0cm}
\begin{abstract}

Heterogeneous Network (HetNet), where Small cell Base Stations (SBSs) are densely deployed to offload traffic from macro Base Stations (BSs), is identified as a key solution to meet the unprecedented mobile traffic demand. 
The high density of SBSs are designed for peak traffic hours and consume an unnecessarily large amount of energy during off-peak time. 
In this paper, we propose a deep reinforcement-learning based SBS activation strategy that activates the optimal subset of SBSs to significantly lower the energy consumption without compromising the quality of service. 
In particular, we formulate the SBS on/off switching problem into a Markov Decision Process that can be solved by Actor Critic (AC) reinforcement learning methods. 
To avoid prohibitively high computational and storage costs of conventional tabular-based approaches, we propose to use deep neural networks to approximate the policy and value functions in the AC approach. 
Moreover, to expedite the training process, we adopt a Deep Deterministic Policy Gradient (DDPG) approach together with a novel action refinement scheme. 
Through extensive numerical simulations, we show that the proposed scheme greatly outperforms the existing methods in terms of both energy efficiency and computational efficiency. 
We also show that the proposed scheme can scale to large system with polynomial complexities in both storage and computation.

\end{abstract}




\section{Introduction}\label{sec:introduction}	

To meet the explosive growth of mobile traffic demand, Small cell Base Stations (SBSs) have been widely deployed to offload the traffic from conventional macro Base Stations (BSs). 
While the dense deployment of SBSs greatly improves the cellular system capacity, it has also made wireless communication networks one of the major sources of the world energy consumption. 
It was reported in \cite{glob_anin} that the Information and Communication Technology (ICT) is responsible for $2\%\sim 10\%$ of the world energy consumption in 2007 and expected to continuously grow further. 
Among the energy consumption of the ICT, more than 80\% is from the Radio Access Network (RAN) \cite{TACT}. 
This is due to the fact that the RAN is deployed to meet the peak traffic load and stays on even when the load is very light. 
As such, energy saving in wireless cellular systems has been seriously investigated for green communications. 
As indicated by \cite{peng_traf}, \cite{mars_opti, kong_rein1, suar_anov, alca_9900}, \cite{moro_impr}, the traffic load of BSs varies drastically at different time of the day. 
According to a real network measurement \cite{peng_traf}, the maximum-to-minimum traffic ratio is larger than five in over $50\%$ of the observed cases, and can be larger than 10 in $30\%$ of the cases. 
This has motivated a surging research interest in base station sleeping/activation scheduling, which activates only necessary BSs to serve the users without a noticeably degradation in quality of service.

Considering the high cost of service migration, service delay, and hardware wear-and-tear, we tend to perform BS on/off operation in a time scale that is much slower than user association in practice. 
Thus, the on/off scheduling decision needs to take into account the spatial-temporal dynamics of network traffic load, user distributions and demands, wireless channel conditions, etc., which vary at faster time scales. 
Previous work on BS on/off scheduling approaches can be broadly categorized into three threads, namely analysis based, optimization based, and learning based methods. 

In particular, analysis based approaches \cite{jiac_reso}, \cite{sung_ener, wuji_traf}, \cite{wuji_base}, \cite{sohy_ener}, \cite{wild_cogn}, \cite{taba_down} focus on the performance analysis assuming a well-captured underlying network and traffic model. 
For mathematical tractability of the analysis, spatial distributions of BSs and users are often assumed to follow Poisson Point Process (PPP) while temporal distributions of traffic arrivals follow Poisson process.  
While they provide systematic insights and convenient metric/parameter evaluation that help the understanding of the studied systems, the analysis results rely heavily on the accuracy and mathematical convenience of the models describing the system dynamics.

On the other hand, \cite{liao_base}, \cite{zhua_ener}, \cite{sonk_base}, \cite{cais_acro} tackle the BS activation problem with optimization approaches. 
These methods can find the optimal or sub-optimal configuration of BS modes under the condition that the stochastic model is perfectly characterized and the needed information is accurate. 
For example, they implicitly assume that the network environment, including user location, service request and channel information, remains unchanged in the considered period. 
However, due to the highly dynamic and stochastic nature in wireless systems, these conditions hardly hold. 
Besides, these methods operate in the time scale of user association (seconds) \cite{zhua_ener}, \cite{cais_acro} or even smaller \cite{liao_base}, which may result in frequent BS mode switching, and ignore the cost of switching on a sleeping BS. 
Moreover, they usually have high computational complexities and need repetitive and intensive computation in each time slot in the scale of seconds.

In contrast, learning based approaches \cite{TACT}, \cite{sake_opti} do not need a well-captured model or any non-causal information. 
Instead, they can learn the environment/model and refine their strategies accordingly. 
The underlying tabular-based structure of these classical learning approaches, however, requires to quantize the continuous state space into a discrete one, making the algorithms unscalable to large systems for two reasons. 
Firstly, the storage space grows exponentially with the state size. 
In our case, typically, the storage space is $S=n_q^{n_b}\cdot 2^{n_b}$, where $n_q$ is the number of quantized values of a state, $n_b$ is the state size (number of SBSs) and $2^{n_b}$ is the number of actions. 
Secondly, the large state-action space makes it extremely difficult to explore and learn in practical time. 
This results in either a long exploration time, which grows exponentially with $n_b$ or a poor sub-optimal strategy.

In this paper, we propose a novel Deep Reinforcement learning based BS Activation alGorithm (DRAG) to efficiently obtain SBS on/off scheduling decisions in large scale HetNets. 
In particular, we first formulate the SBS mode scheduling as a Markov Decision Process (MDP) that operates in large time scale and utilizes statistics rather than snapshots of user activity information.  
To solve the MDP, we combine the latest advance in deep learning and Actor-Critic (AC) Reinforcement Learning (RL). 
Specifically, we approximate both the Actor (policy function) and the Critic (value function) of the AC-RL framework with Deep Neural Networks (DNN) and train the system with the Deep Deterministic Policy Gradient (DDPG) algorithm \cite{DDPG}. 
The proposed approach DRAG does not depend on the distributions of channel, traffic demand, user location, or the underlying resource allocation/scheduling mechanisms. 
In fact, it is purely data-driven and completely model-free. 
Moreover, with the excellent generalization capability and approximation capacity of the DNNs, it avoids the problems of large-storage and slow learning of conventional tabular-based RL methods. 
Being able to learn online, it can effectively exploit the traffic pattern and adapt to the varying environment. 

We highlight our contributions as follows. 
\begin{enumerate}

\item We propose a DRL approach to solve the BS activation scheduling problem and show that it can successfully apply to HetNets with many SBSs and continuous state of traffic demand.

\item We explicitly exploit the spatial and temporal correlation of data traffic arrivals to jointly predict the traffic arrivals at all SBSs with a DNN.

\item We propose an $\epsilon$-cost-greedy action refinement procedure to assist exploration. 
The proposed procedure overcomes the inefficiency of the conventional randomness-based exploration approaches and significantly accelerates the learning process. 

\end{enumerate}

The remainder of this paper is organized as follows. 
We introduce the system model and describe the problem formulation in Section \ref{sec:system}. 
Deep reinforcement learning solution is described in Section \ref{sec:DRL}. 
Numerical results are presented in Section \ref{sec:results}. 
Finally, this paper is concluded in Section \ref{sec:conslusion}.

\section{System Model}\label{sec:system}


\subsection{Network Scenario}\label{subsec:network}

\begin{figure}[h!]
\centering
\includegraphics[width=2.8in]{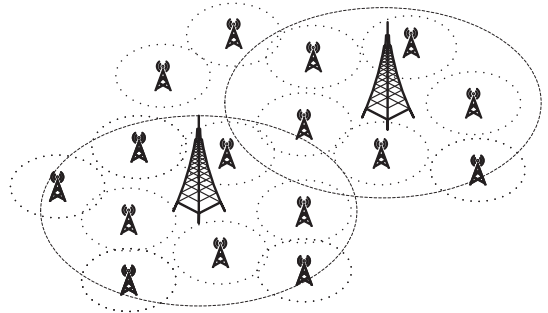}
\caption{A heterogeneous network with two MBSs and many SBSs. } \label{fig:hetnet}
\end{figure}

We consider a large-scale Heterogeneous Network (HetNet) consisting of a number of Macro Base Stations (MBSs) and Small Base Stations (SBS), as depicted in Fig. \ref{fig:hetnet}. 
Denote the set and number of MBSs by $\mathcal{B}_m$ and $B_m$, respectively. 
Likewise, denote the set and number of SBSs by $\mathcal{B}_s$ and $B_s$, respectively. 
Let $\mathcal{B} = \mathcal{B}_m \cup \mathcal{B}_s$ and $B = B_m + B_s $. 
Suppose that the SBSs are deployed at traffic hot spots to handle peak traffic loads, and that they do not have significant coverage overlap. 
This can be achieved by dedicated design on SBS locations in traffic hot-spots and configuration on the SBSs' functioning parameters, e.g. transmit power, antenna direction etc.. 
Suppose that the SBSs all reuse the MBS's frequency resource. 
Similar to many existing 4G systems, we assume that the SBSs reuse the MBSs' frequency spectrum. 
The interference between the MBS and SBSs can be suppressed by advanced MIMO and interference mitigation techniques. 
We assume that a MBS has larger capacity than a SBS due to its higher antenna, multiple sectors, larger transmit power, clearer channel path and more antennas \cite{deru_powe}.

Suppose that each SBS is equipped with a low-power traffic monitor \cite{ashr_slee} and a remote controller that is connected to the MBS and can change the SBS's operation mode. 
In particular, the low-power traffic monitor can sense the traffic load, i.e., user data Arrival Rate (AR), even when the BS is turned off. 
On the other hand, the remote controller enables the MBS to turn on the SBS when it is off. 

We discretize the time into time slots. 
Each time slot has a span of half an hour, during which the modes of the SBSs remain unchanged. 
Let $\bm{v}^t$ be the indicator variable vector denoting the on/off modes of the BSs in time slot $t$, with $v_i^t=1$ indicating that SBS $i$ is on/active and $v_i^t=0$ otherwise.

\subsection{Energy Consumption}\label{subsec:power}

The energy consumption of a typical base station consists of two parts: a constant energy consumption that is irrelevant to the BS traffic load and a load-dependent energy consumption that is proportional to the traffic load \cite{TACT, sonk_base}. 
The constant energy consumption comes from the peripheral equipments (power supply and air conditioner) and the always-on components of the communication subsystem (transceiver), while the load-dependent energy comes from the power amplifier \cite{deru_powe}. 
Overall, we denote the energy consumption of a general BS $i$ at time slot $t$ by
\begin{equation}\label{eqn:power}
\begin{array}{rl}
p_i^t=P^c_i + \rho_i^t\cdot P^l_i,
\end{array}
\end{equation}
where $P^c_i$ is the constant energy, $P^l_i$ is the load-dependent energy and $\rho_i^t$ is the traffic load of BS $i$ at time slot $t$. 
In view of the variation of traffic demands, we investigate the on/off scheduling of SBSs, assuming that the MBS is always on. 
Note that our approach is model-free, data-driven and does not need the explicit knowledge of the energy consumption model in Eqn. (\ref{eqn:power}) in practice.

\subsection{Traffic Pattern}\label{subsec:traffic}

Denote the traffic arrival rate (bps) at SBS $i$ in time slot $t$ by $\lambda_{i}^t$. 
The system load of SBS $i$, $\rho_i^t$, is a function of both $\lambda_i^t$ and the service rate the active BSs can provide, which is affected by $\bm{v}^t$. 
We assume that the time slot is long enough, so that the traffic load of one slot does not carry on to the next slot. 
Thus, the system load is
\begin{equation}\label{eqn:load}
\begin{array}{rl}
\rho_i^t = f_l(\lambda_{i}^t,v_i^t), 
\end{array}
\end{equation}
where $f_l(\cdot)$ is a function mapping the arrival rates and SBS modes to the load. 
Due to complicated dynamics in user behavior, traffic demand per connection, service elasticity, etc., accurate modeling of $f_l$ is hard in general. 
This motivates us to adopt a data-driven approach in the next section.

According to the measurement results in real networks \cite{peng_traf}, \cite{mars_opti, suar_anov, alca_9900}, the traffic arrival of a BS fluctuates dramatically across a day. 
Moreover, the fluctuation shows strong repetitive pattern every day or workday, which indicates strong correlation in the ARs across time. 
Inspired by this correlation, we seek to predict the short term AR with historic AR data, which will be detailed in sub-section \ref{subsec:arp}.


\subsection{Problem Formulation}

\begin{figure}[h!]
\centering
\includegraphics[width=2.0in]{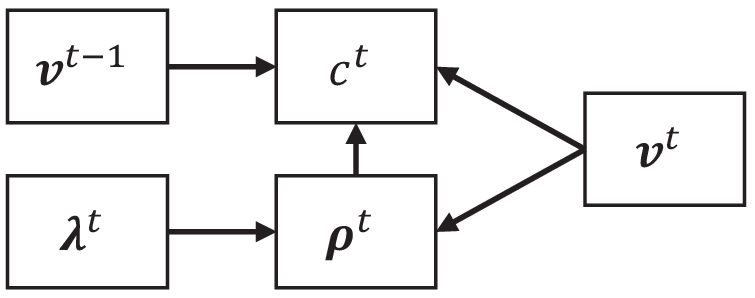}
\caption{Dependency of the cost. } \label{fig:logic_flow}
\end{figure}

Before introducing the problem formulation, we first describe the cost metrics being considered. 
In addition to the energy consumption of the network, we also consider the Quality of Service (QoS) degradation cost and mode switching cost \cite{wuji_traf}. 
The QoS degradation cost captures the negative effect of turning off SBSs on QoS and is adopted to prevent turning off SBSs too aggressively. 
Meanwhile, the switching cost, which captures the power surge and measures the harmful effect to the hardware, is incorporated to prevent changing SBS modes too frequently. 
Specifically, the QoS degradation cost is denoted as 
\begin{equation}\label{eqn:delay_cost}
\begin{array}{lrrl}
c_d^t (\bm{\lambda}^t, \bm{v}^t) = \beta_d \cdot f_d(\bm{\lambda}^t,\bm{v}^t),
\end{array}
\end{equation}
where $\beta_d$ is the penalty factor and $f_d(\bm{\lambda}^t,\bm{v}^t)$ is the mapping from ARs and SBS modes to QoS degradation. 
A special case of the QoS degradation cost is the service delay cost adopted in \cite{TACT}. 
In our data-driven approach, the QoS degradation function does not need to be explicitly modeled. 
Instead, it is measured at the end of each time slot. 
On the other hand, the switching cost is denoted by 
\begin{equation}\label{eqn:switching_cost}
\begin{array}{lrrl}
c_s^t (\bm{v}^{t-1},\bm{v}^t) = \beta_s \cdot \sum\limits_{i\in \mathcal{B}_s} (v^t_i-v^{t-1}_{i})^+ ,
\end{array}
\end{equation}
where $\beta_s$ is the penalty factor of unit $W$, and the function $(\cdot)^+$ is defined as
\begin{equation}\label{eqn:relu}
(x)^+=\left\{
\begin{aligned}
x & \text{, if } x\geq 0, \\
0 & \text{, if } x<0. \\
\end{aligned}
\right.
\end{equation}
As such, the total cost of the system in time slot $t$ is given by 
\begin{equation}\label{eqn:total_cost}
\begin{array}{lrrl}
c^t = \sum\limits_{i\in \mathcal{B} } p_i^t + c_d^t + c_s^t. 
\end{array}
\end{equation}
For better illustration, we show the dependency of the cost in Fig. \ref{fig:logic_flow}.

We are now ready to formulate the total cost minimization problem in HetNet as a Markov Decision Process (MDP). 
An MDP is defined by a 5-tuple $\mathcal{M}=(\mathcal{S},\mathcal{V},P,C,\gamma)$, where $\mathcal{S}$ is the state space, $\mathcal{V}$ is the action space, $P$ is the state transition probability function, $C$ is the cost function and $\gamma\in[0,1]$ is the discount factor. 
At the beginning of each time slot $t$, the system controller/agent sees the state $\bm{s}^t$ and correspondingly selects an action $\bm{v}^t$ to execute. 
Then, the agent receives a cost $c^t$ at the end of time slot $t$ and sees a new state $\bm{s}^{t+1}$. 
We introduce the MDP design as follows. 
\subsubsection{State}
At each time slot $t$, the system state $\bm{s}^t\in \mathcal{S}$ is defined by $\bm{s}^t=(\bm{\lambda}^{t}, \bm{v}^{t-1})$. 
In practice, the ARs $\bm{\lambda}^{t}$ is unavailable at the beginning of time slot $t$ when the decision is to be made. 
Alternatively, we use the predicted value $\hat{\bm{\lambda}}^{t}$, which is obtained with reasonable accuracy by our proposed AR prediction module. 

\subsubsection{Action and Transition Function}
The action at time slot $t$ is the SBS on-off decision, i.e. $\bm{v}^t\in \mathcal{A}$. 
The state transition probability function $P\left( \bm{s}^{t+1}|\bm{s}^t,\bm{v}^t \right)$ represents the distribution of the next state $\bm{s}^{t+1}$ given the current state $\bm{s}^t$ and action $\bm{v}^t$.

\subsubsection{Cost Function}
The cost function $C(\bm{s}^t,\bm{v}^t)$ captures the immediate cost at time slot $t$ when the system transits from state $\bm{s}^t$ to $\bm{s}^{t+1}$ due to $\bm{v}^t$, and is defined by Eqn. (\ref{eqn:total_cost}), i.e. 
\begin{equation}\label{eqn:cost_fun}
\begin{array}{lrrl}
c^t = C(\bm{s}^t,\bm{v}^{t}) = \sum\limits_{i\in \mathcal{B} } p_i^t(\bm{\lambda}^t, \bm{v}^t) + c_d^t(\bm{\lambda}^t, \bm{v}^t) + c_s^t(\bm{v}^{t-1}, \bm{v}^{t}). 
\end{array}
\end{equation}

At each time slot, we aim to find an action that minimizes the long-term cost. 
To evaluate the long-term cost of an action given a state, we consider the state-action-value function $Q(\bm{s}^t, \bm{v})$, which is defined as the expected cost at state $\bm{s}^t$ when action $\bm{v}$ is taken, i.e.,
\begin{equation}\label{eqn:q-function}
\begin{array}{ll}
Q(\bm{s}^t, \bm{v})
 = C(\bm{s}^t,\bm{v}) + \gamma\cdot \min\limits_{\bm{v}^{\prime}} \mathbb{E} \big[  Q\left( \bm{s}^{t+1}, \bm{v}^{\prime} \right) \big].\\
\end{array}
\end{equation}
The problem then becomes 
\begin{equation}\label{eqn:prob}
\begin{array}{ll}
\min\limits_{\bm{v}} & Q(\bm{s}^t, \bm{v} )\\
s.t. & v_i \in \{0,1\},\forall i.
\end{array}
\end{equation}
Our objective is to find the action that solves problem (\ref{eqn:prob}) given the system state $\bm{s}^t$, i.e. $\bm{v}^t = \arg \min\limits_{\bm{v}} Q(\bm{s}^t, \bm{v} )$.



\section{Deep Reinforcement-Learning Based Solution}\label{sec:DRL}

An MDP is generally tackled by Dynamic Programming (DP) \cite{sutt_rein} or Reinforcement Learning (RL) approaches \cite{igro_effi, jpet_natu, q_learn, SARSA}. 
The DP approaches heavily rely on a well-defined mathematical model, e.g., the transition probabilities, of the underlying system. 
On the other hand, the RL methods do not require any assumptions on the model. 
Instead, they learn the model through interacting with the environment. 

\begin{figure}[h!]
\centering                                                         
\includegraphics[width=1.5in]{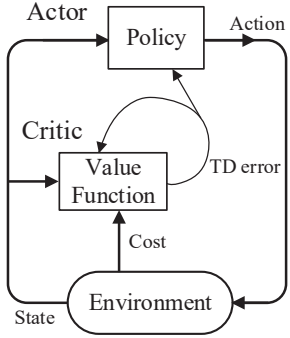}
\caption{The Actor-Critic reinforcement learning framework.}\label{fig:AC}
\end{figure}

In this paper, we adopt the Actor-Critic (AC) \cite{AC} RL framework, as shown in Fig. \ref{fig:AC}, to solve problem (\ref{eqn:prob}). 
The AC-RL approach inherits the advantages of both the value-based \cite{q_learn, SARSA}. 
Specifically, the Critic refines the value function with the Time Difference (TD)-error and then criticizes the policy with the TD-error, guiding the Actor to produce an action with low cost value. 
The Actor, which is usually a parameterized policy, generates an action (distribution) given the state and updates the policy with the TD-error from the Critic. 
Conventionally, the Actor uses a Boltzmann \cite{TACT} or Gaussian \cite{weiy_user} probability distribution as the policy while the Critic utilizes a state-action value table as the value function. 
Due to the page limit, readers are referred to the references for the details of the AC-RL method \cite{TACT},\cite{weiy_user}.

Classical AC approaches, while performing well in small scale problems, often suffer the large storage space and inefficient learning in large-scale problems. 
The large storage problem is due to the fact that the tabular-based Critic needs to quantize the state and action spaces, which results in an exponentially growing storage with respect to the number of state and action variables. 
On the other hand, the inefficient learning arises from the diminishing efficiency of randomness-based exploration procedure due to the curse of dimensionality and the absence of generalization among neighboring state-action pairs in tabular-based AC approaches. 
In fact, most practical problems are too large to learn all action values in all states separately \cite{DDQN}. 
In our problem where there are up to tens of SBSs, the classical AC-RL methods hardly produce satisfactory performance, as can be seen in our simulations.

One way to address the abovementioned issue is to utilize continuous parameterized functions to approximate the policy function and the value function \cite{weiy_user} instead of using tabular based approaches. 
In particular, we adopt the Deep Neural Network (DNN) to approximate the continuous functions \cite{DDQN}, \cite{DQN}, \cite{maoh_neur}. 
DNN has been identified as a universal function approximator that can approximate any function mapping, possibly with multiple inputs and multiple outputs, to arbitrary precision given enough neural units \cite{horn_mult}. 
We formally denote a parameterized DNN approximator as $f(x|\theta)$, where $f$ is the output of the DNN, $x$ is the input and $\theta$ is the network parameters, including the weights and biases. 
Besides, DNN is also used for arrival rate prediction in our work.

\begin{figure}[h!]
\centering
\includegraphics[width=3.0in]{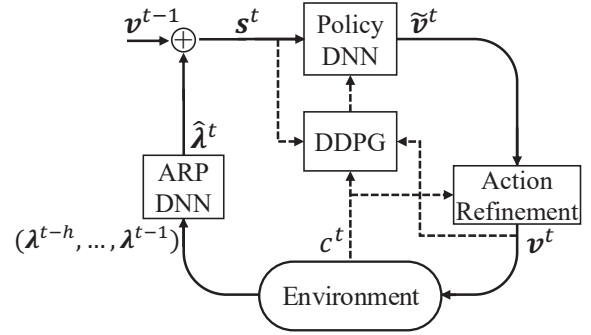}
\caption{The whole schematic of the proposed algorithm. } \label{fig:ACDQN}
\end{figure}

The structure of our approach is depicted in Fig. \ref{fig:ACDQN}. 
At the beginning of time slot $t$, we first predict the AR during the time slot with truncated historic ARs by the AR Prediction DNN (ARP DNN). 
The predicted AR $\hat{ \bm{\lambda}}^t$, together with $\bm{v}^{t-1}$, constitutes the state $\bm{s}^t$, which is then fed to the policy DNN (Actor) to generate an action $\tilde{ \bm{v}}^t$. 
Before executing the action, we adopt an action-refinement module to enhance the exploration. 
At the end of time slot $t$, the agent receives a cost $c^t$. 
We note that the Critic, whose role is played by the value DNN, and its training are contained in the DDPG module. 
The details of the individual modules will be elaborated in the following subsections.

\subsection{Arrival Rate Predictor}\label{subsec:arp}

In this sub-section, we describe the details of the ARP DNN. 

In practice, traffic arrival rates are strongly correlated both across consecutive time slots and between neighboring BSs. 
We exploit such correlation to forecast the AR for time slot $t$ based on the truncated historic AR data 
\begin{equation}\label{eqn:har}
\begin{array}{lrrl}
\bar{\bm{\lambda}}^{t} = (\bm{\lambda}^{t-h},...,\bm{\lambda}^{t-1}),
\end{array}
\end{equation} 
where $h$ is the history length. 
Specifically, the ARP DNN takes $\bar{\bm{\lambda}}^{t}$ as input and outputs the predicted AR $\hat{\bm{\lambda}}^t $, and is denoted by
\begin{equation}\label{eqn:arpn}
\begin{array}{lrrl}
\hat{\bm{\lambda}}^t = A(\bar{\bm{\lambda}}^{t}| \theta_A),
\end{array}
\end{equation}
where $\theta_A$ denotes the trainable parameters of the network. 
Note that $\hat{\bm{\lambda}}^t$ is a vector containing ARs at all BSs.

The data samples for APR DNN training are obtained by storing the historic and real AR data pairs $(\bar{\bm{\lambda}}^{t}, \bm{\lambda}^t)$ in a replay memory $R_A$ every time the real AR is revealed at the end of a time slot. 
To train the DNN, we aim to minimize the prediction error/loss, 
\begin{equation}\label{eqn:arpp_error}
\begin{array}{lrrl}
L_{A} = \mathbb{E} \left[ \left( \hat{\bm{\lambda}}^i - \bm{\lambda}^{i} \right)^2 \right], 
\end{array}
\end{equation}
where $\hat{\bm{\lambda}}^i = A(\bar{\bm{\lambda}}^{i}|\theta_{A} )$, $(\bar{\bm{\lambda}}^{i}, \bm{\lambda}^{i})\in R_A$, with gradient-descent method. 
The gradient of $L_{A}$ with respect to the network parameters $\theta_{A}$ is
\begin{equation}\label{eqn:arpp_grad}
\begin{array}{lrrl}
\bigtriangledown_{\theta_{A}} L_{A} = \mathbb{E} \left[ 2\left( \hat{\bm{\lambda}}^i - \bm{\lambda}^{i} \right) \cdot \bigtriangledown_{\theta_{A}} A(\bar{\bm{\lambda}}^{i}|\theta_{A}) \right].
\end{array}
\end{equation}
In practice, at each training step, $\theta_{A}$ is updated with a mini-batch $(\bar{\bm{\lambda}}^{i}, \bm{\lambda}^{i}), i=1,...,N,$ that are randomly fetched from the replay memory $R_A$, i.e.,
\begin{equation}\label{eqn:arpp_update}
\begin{array}{lrrl}
\theta_{A} = \theta_{A} - \frac{\alpha_{A}}{N} \sum\limits_{i=1}^{N} 2 \left( A\left( \bar{\bm{\lambda}}^{i}|\theta_{A} \right) - \bm{\lambda}^{i} \right) \cdot \bigtriangledown_{\theta_{A}} A\left( \bar{\bm{\lambda}}^{i}|\theta_{A} \right), 
\end{array}
\end{equation}
where $\alpha_{A}$ is the learning rate.

\subsection{The Policy and Value DNNs}\label{subsec:ac}

Conventionally, the Actor network (i.e. the policy DNN) of a DRL system outputs a probability distribution of possible actions \cite{DQN, maoh_neur}. 
However, this is only feasible when the number of possible actions is small. 
In our case, the number of actions $2^{B_s}$ can be huge for large $B_s$. 
Outputting a probability distribution would need a huge number of output neurons. 
To overcome the issue, our policy DNN directly outputs a single deterministic action vector to represent the SBS mode selection. 

Formally, we represent the the Actor network (policy DNN) as
\begin{equation}\label{eqn:pdnn}
\begin{array}{lrrl}
\tilde{\bm{v}}^t = \pi \left( \bm{s}^t | \theta_{\pi} \right), \\
\end{array}
\end{equation}
where the input $\bm{s}^t$ is the state, the output $\tilde{\bm{v}}^t$ is the action, and $\theta_{\pi}$ denotes the parameters of the policy DNN. 
We adopt the modified $tanh(x)$ function as the activation of the output layer to confine the output values to $[0,1]$. 
Specifically, we use $\frac{tanh(x+2)+1}{2}$ to encourage the DNN to output 1 in initial time slots, in order to avoid severe traffic congestion in the MBSs. 
Note that the output of the policy DNN $\tilde{\bm{v}}^t$ is a vector of continuous values, which is not a valid SBS mode selection decision. 
To convert it into a valid decision vector $\bm{v}^t$, we will refine $\tilde{\bm{v}}^t$ in sub-section \ref{subsec:action_refine}. 
The network parameters $\theta_{\pi}$ are optimized by gradient descent with the Back-Propagation (BP) algorithm. 
The details are deferred to sub-section \ref{subsec:ddpg}.


The Critic network, namely the value DNN, approximates the Q-value of a given state-action pair. 
It is represented as
\begin{equation}\label{eqn:vdnn}
\begin{array}{lrrl}
q^t = Q \left( \bm{s}^t, \bm{v}^t | \theta_Q \right), \\
\end{array}
\end{equation}
where the inputs $\bm{s}^t, ~\bm{v}^t$ are the state and action, respectively, and $\theta_Q$ denotes the trainable parameters of the value DNN. 
The value DNN takes the state $\bm{s}^t$ and action $\bm{v}^t$ as inputs, and outputs the corresponding estimated Q-value $q^t$. 
Similar to the policy DNN, the network parameters $\theta_Q$ are also trained with the BP algorithm. 
The details are deferred to sub-section \ref{subsec:ddpg}. 

We note that the value DNN does not directly participate in the action generation. 
Instead, it serves as a guidance for training the policy DNN in an asynchronous way.

\subsection{Action Refinement}\label{subsec:action_refine}

\begin{figure}[h!]
\centering
\includegraphics[width=3.0in]{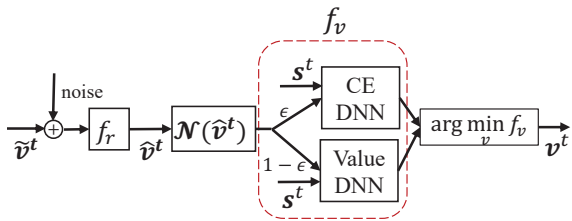}
\caption{Details of the action refinement module. } \label{fig:action_refine}
\end{figure}

Exploration is one of the key ingredients to the success of a DRL system. 
As mentioned in the last subsection, our policy DNN deterministically outputs a single action to reduce complexity, which results in little exploration. 
To compensate for the lack of exploration, a commonly used solution is to add noise to the outputted action $\tilde{\bm{v}}^t$ \cite{DDPG} or to the Actor network parameters $\theta_{\pi}$ \cite{plap_para}. 
We adopt the first method here as part of our action refinement. 
The details of the whole action refinement are depicted in Fig. \ref{fig:action_refine}. 
To begin with, we define the proto-action as
\begin{equation}\label{eqn:noi_exploration}
\begin{array}{lrrl}
\hat{\bm{v}}^t = f_r\left( \tilde{\bm{v}}^t + \mathbb{N}(\sigma_n(t)) \right),
\end{array}
\end{equation}
where $\sigma_n(t)$ is the decaying noise intensity and $f_r(\cdot)$ is a function that maps the continuous  action to a binary one. 
However, due to the Curse of Dimensionality, the random noise-based approach can only provide coarse exploration  in a high dimensional action space. 
An intuitive example is that in our case, the optimal action in some light-traffic time slot is the all-zero vector, i.e. all SBSs should be switched off. 
However, the probability to generate such specific action with random noise-based exploration diminishes exponentially with $B_s$.

To this end, we need additional procedure to aid exploration. 
A Q-value guided exploration called the Wolpertinger policy has been proposed in \cite{action_refine}. 
Before executing the action, the Wolpertinger policy first refines the action as follows.
\begin{equation}\label{eqn:action_refine}
\begin{array}{lrrl}
\bm{v}^t = \arg \min\limits_{\bm{v}\in \mathcal{A}_k} Q \left( \bm{s}^{t}, \bm{v} \right),
\end{array}
\end{equation}
where $\mathcal{A}_k$ is a set of $k$ closest actions of $\tilde{\bm{v}}^t$. 
However, in practice, the Q-value estimation takes a long time to converge due to its unsupervised training. 
Alternatively, we notice that the cost-greedy action usually is near optimal and provides good direction for action refinement. 
To guide the action refinement, we adopt a Cost-Estimation DNN (CEN) to estimate the cost, i.e.
\begin{equation}\label{eqn:cost_esti}
\begin{array}{lrrl}
\hat{c}^t = \tilde{C}\left( \hat{\bm{\lambda}}^t, \bm{v}^{t-1}, \bm{v} | \theta_{\tilde{C}} \right),
\end{array}
\end{equation}
where $\hat{\bm{\lambda}}^t, \bm{v}^{t-1}, \bm{v}$ are the inputs and $\theta_{\tilde{C}}$ denotes its parameters. 
With the CEN, we propose a neighborhood based action refinement method. 
Specifically, we explore the neighborhood of $\hat{\bm{v}}^t$ and find the action with the minimum estimated cost,
\begin{equation}\label{eqn:action_refine}
\begin{array}{lrrl}
\bm{v}^t = \arg \min\limits_{\bm{v}\in \mathcal{N} (\hat{\bm{v}}^t)} \tilde{C}\left( \hat{\bm{\lambda}}^t, \bm{v}^{t-1}, \bm{v} | \theta_{\tilde{C}} \right),
\end{array}
\end{equation}
where the neighborhood of $\hat{\bm{v}}^t$ is defined as $\mathcal{N} (\hat{\bm{v}}^t) = \{\bm{v}| ~ ||\bm{v}-\hat{\bm{v}}^t||^2 \leq D \}$ and $D$ is a tunable distance parameter. 
We refer to this procedure as cost-greedy exploration.

The main advantage of the cost based refinement method over the Q-value based counterpart is that the CEN can be trained with supervised training, which is much easier and faster than the unsupervised training in the Q-value estimation. 
In fact, similar to the training of APR DNN, we can train the CEN with the sequentially revealed $\left( (\bm{\lambda}^t, \bm{v}^{t-1}, \bm{v}^t), c^t \right)$ information. 
To do that, we store the $\left( (\bm{\lambda}^i, \bm{v}^{i-1}, \bm{v}^i), c^i \right)$ tuples in a replay memory $R_{\tilde{C}}$ every time the cost information is revealed at the end of a time slot. 
Then, the network is trained to minimize the cost estimation loss
\begin{equation}\label{eqn:lmp_error}
\begin{array}{lrrl}
L_{\tilde{C}} = \mathbb{E} \left[ \left( \hat{c}^i - c^{i} \right)^2 \right], 
\end{array}
\end{equation}
where $\hat{c}^i = \tilde{C}\left( \bm{\lambda}^i, \bm{v}^{i-1}, \bm{v}^i | \theta_{\tilde{C}} \right) $.
The gradient of $L_{\tilde{C}}$ with respect to $\theta_{\tilde{C}}$ is
\begin{equation}\label{eqn:lmp_grad}
\begin{array}{lrrl}
\bigtriangledown_{\theta_{\tilde{C}}} L_{\tilde{C}} = \mathbb{E} \left[ 2\left( \hat{c}^i - c^{i} \right) \cdot \bigtriangledown_{\theta_{\tilde{C}}} \tilde{C}(\bm{\lambda}^i, \bm{v}^{i-1}, \bm{v}^t|\theta_{\tilde{C}}) \right]. 
\end{array}
\end{equation}
At each training step, $\theta_{\tilde{C}}$ is updated with a mini-batch $ (\bm{\lambda}^i, \bm{v}^{i-1}, \bm{v}^i, c^i), i=1,...,N,$ that are randomly fetched from the replay memory, i.e.,
\begin{equation}\label{eqn:lmp_update}
\begin{array}{lrrl}
\theta_{\tilde{C}} = \theta_{\tilde{C}} - \frac{\alpha_{\tilde{C}}}{N} \sum\limits_{i=1}^{N} 2\left( \hat{c}^i - c^{i} \right) \cdot \bigtriangledown_{\theta_{\tilde{C}}} \tilde{C} \left( \bm{\lambda}^i,\bm{v}^{i-1}, \bm{v}^i|\theta_{\tilde{C}} \right), 
\end{array}
\end{equation}
where $\alpha_{\tilde{C}}$ is the learning rate.

We are aware that the cost-greedy refinement method in general does not obtain the same optimal refinement direction as the Q-value based method does. 
To exploit the advantages of both methods, we use a hybrid action refinement (exploration) procedure, 
\begin{equation}\label{eqn:action_refine1}
\begin{array}{lrrl}
 \bm{v}^t = \arg \min\limits_{\bm{v}\in \mathcal{N} (\hat{\bm{v}}^t)} f_v(\bm{s}^t,\bm{v}),
\end{array}
\end{equation}
where
\begin{equation}\label{eqn:action_refine2}
f_v(\bm{s}^t,\bm{v}) = \left\{
\begin{aligned}
\tilde{C}\left( \hat{\bm{\lambda}}^t, \bm{v}^{t-1}, \bm{v} \right)  \text{, if } r \leq \epsilon(t), \\
Q^t\left( \bm{s}^t,\bm{v} \right)  \text{, if } r > \epsilon(t). \\
\end{aligned}
\right.
\end{equation}
In Eqn. (\ref{eqn:action_refine2}), $r$ is a random number following a uniform distribution in $[0,1]$ and $\epsilon(t)$ is a decreasing threshold that controls the tendency of selecting the refinement metric.

\subsection{The Training Algorithm for AC Networks}\label{subsec:ddpg}

In contrast to the training of the APR DNN and CEN, neither the policy DNN nor the value DNN can be trained in a supervised fashion, since the ground truth is not known beforehand. 
Here, we adopt the Deep Deterministic Policy Gradient (DDPG) \cite{DDPG} training framework, which is recently proposed to train Actor and Critic networks with improved stability. 

\begin{figure}[h!]
\centering
\includegraphics[width=2.0in]{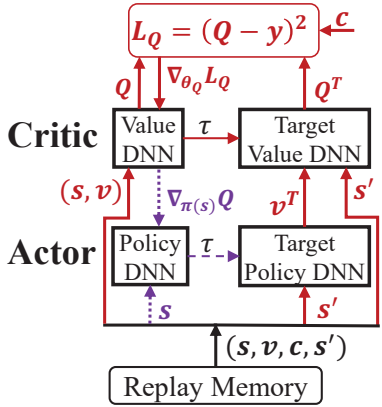}
\caption{The DDPG training algorithm of the AC networks. The red solid lines correspond to the Critic network training, while the blue dashed lines correspond to the Actor network training. } \label{fig:DDPG}
\end{figure}

DDPG is an Actor-Critic framework based algorithm, in which the policy and value function are both approximated by DNNs. 
As depicted in Fig. \ref{fig:DDPG}, in addition to the policy and value DNNs, the DDPG framework has target policy and target value DNNs that have exactly the same structure as the policy and value DNNs. 
The target networks, denoted by $\pi(\bm{s}^t|\theta_{\pi}^T)$ and $Q(\bm{s}^t, \bm{v}^t|\theta_Q^T)$, are only lagged versions of the original ones. 
The target networks are used to generate target/objective Q-values for training the original value DNN, as will be described in \ref{subsubsec:ucn}. 
Furthermore, DDPG utilizes a replay memory $R_{ac}$ to store the experiences for network training. 
The target networks provide a stable target value source. 
Meanwhile, the random experience fetched from the replay memory breaks up the correlation among the experiences in a mini-batch. 
As shown in \cite{DQN}, these two blocks significantly improve the training stability.

Before introducing the network training process, we first describe how the AC networks work. 
Given the system state $\bm{s}^t=(\hat{\bm{\lambda}}^t, \bm{v}^{t-1})$ at time slot $t$, the Actor network generates an action $\tilde{\bm{v}}^t$, which further yields $\bm{v}^t$ after action refinement. 
After operating for a time slot with $\bm{v}^t$, the system/agent receives a cost $c^t$ and enters a new state $\bm{s}^{t\prime}=(\hat{\bm{\lambda}}^{t+1}, \bm{v}^{t})$. 
The agent then stores the experience tuple $(\bm{s}^t,\bm{v}^t,c^t,\bm{s}^{t\prime})$ in the replay memory $R_{ac}$, which will be used in the training process as described in the following.

\subsubsection{Critic Network Training}\label{subsubsec:ucn}

The Critic network parameters $\theta_Q$ is trained in a semi-supervised way by treating the values produced by the target networks as the (approximately) true objective. 
Specifically, in the Critic network, we seek to minimize the loss, defined as follows, with respect to the experiences, 
\begin{equation}\label{eqn:q_cost}
\begin{array}{lrrl}
L_Q(\theta_Q) = \mathbb{E}\left[ \left( y -  Q\left(\bm{s},\bm{v}|\theta_Q \right) \right)^2 \right], 
\end{array}
\end{equation}
where
\begin{equation}\label{eqn:tq}
\begin{array}{lrrl}
y = c + \gamma \cdot Q \left( \bm{s}^{\prime}, \pi\left(\bm{s}^{\prime}|\theta_{\pi}^T \right)|\theta_Q^T \right) 
\end{array}
\end{equation}
is the objective/target $Q$-value and $\bm{s}^{\prime}$ is the next state after taking action $\bm{v}$. 
Notice that the objective $Q$-values are obtained through the target networks, i.e. networks with parameters $\theta_{\pi}^T$ and $\theta_Q^T$. 

The gradient of $L_Q(\theta_Q)$ with respect to $\theta_Q$ is computed as 
\begin{equation}\label{eqn:q_grad}
\begin{array}{lrrl}
\bigtriangledown_{\theta_Q} L_Q(\bm{s},\bm{v}) = \mathbb{E}\left[ 2\cdot \left( y - Q\left( \bm{s},\bm{v}|\theta_Q \right) \right) \cdot \bigtriangledown_{\theta_Q} Q\left( \bm{s},\bm{v} \right) \right],
\end{array}
\end{equation}
where $\bigtriangledown_{\theta_Q} Q\left( \bm{s},\bm{v} \right)$ is computed by the chain-rule from the output back to each specific parameters in $\theta_Q$. 
We note that in Eqn. (\ref{eqn:q_grad}), $ y - Q(\bm{s},\bm{v}|\theta_Q) $ is actually the TD-error. 
In practice, the network is trained by stochastic gradient descent. 
That is, at each training step, we update $\theta_Q$ with a mini-batch experiences $(\bm{s}^i,\bm{v}^i,c^i,\bm{s}^{i\prime}), i=1,...,N,$ that are randomly sampled from the replay memory $R_{ac}$, 
\begin{equation}\label{eqn:cnp_update}
\begin{array}{lrrl}
\theta_Q = \theta_Q -  \frac{\alpha_{Q}}{N} \sum\limits_{i=1}^{N} 2\cdot \left( y^i - Q(\bm{s}^i,\bm{v}^i|\theta_Q) \right) \cdot \bigtriangledown_{\theta_Q} Q\left( \bm{s}^i,\bm{v}^i \right), 
\end{array}
\end{equation}
where $y^i = c^i + \gamma \cdot Q\left( \bm{s}^{\prime}, \pi\left( \bm{s}^{\prime}|\theta_{\pi}^T \right)|\theta_Q^T \right)$ and $\alpha_{Q}$ is the learning rate of the Critic network.

\subsubsection{Actor Network Training}

In the Actor network, the objective is to minimize the loss
\begin{equation}\label{eqn:a_cost}
\begin{array}{lrrl}
L_{\pi}(\theta_{\pi}) = \mathbb{E}\left[ \left( \bm{v}^* - \bm{v} \right)^2 \right],
\end{array}
\end{equation}
where $\bm{v}= \pi(\bm{s}|\theta_{\pi})$ and $ \bm{v}^* $ denotes the optimal action. 
However, we cannot minimize this cost by gradient descent in a similar way as the Critic network training. 
This is because $\bm{v}^*$ can only be found by exhaustively evaluating the Critic network $\bm{v}^* \approx \arg\min_{\bm{v}} Q(\bm{s},\bm{v}|\theta_Q)$ assuming that the value DNN is accurate. 
To this end, instead of minimizing the loss, we update the Actor network parameters $\theta_{\pi}$ towards minimizing the Q-value of the outputted action with the deterministic policy gradient \cite{DPG}. 
To do that, we need the gradient form the Critic network with respect to the Actor network's output action $\tilde{ \bm{v}}=\pi(\bm{s}|\theta_{\pi})$, i.e. $ \bigtriangledown_{\tilde{ \bm{v}}} Q(\bm{s},\tilde{ \bm{v}})$. 
The complete gradient is
\begin{equation}\label{eqn:a_grad}
\begin{array}{lrrl}
\bigtriangledown_{\theta_{\pi}} Q = \bigtriangledown_{\bm{v}} Q(\bm{s},\bm{v}|\theta_Q)|_{\bm{v}=\pi(\bm{s}|\theta_{\pi})} \cdot \bigtriangledown_{\theta_{\pi}} \pi(\bm{s}),
\end{array}
\end{equation}
where $\bigtriangledown_{\theta_{\pi}} \pi(\bm{s})$ is computed by the chain-rule. 
It is proved in \cite{DPG} that the stochastic policy gradient, which is the gradient of the policy's performance, is equivalent to the empirical deterministic policy gradient, i.e.,
\begin{equation}\label{eqn:a_grad1}
\begin{array}{lrrl}
\bigtriangledown_{\theta_{\pi}} Q^* 
\approx \mathbb{E}_{\pi}\left[ \bigtriangledown_{\bm{v}} Q(\bm{s},\bm{v}|\theta_Q)|_{\bm{v}=\pi(\bm{s}|\theta_{\pi})} \cdot \bigtriangledown_{\theta_{\pi}} \pi(\bm{s}) \right].
\end{array}
\end{equation}
At each training step, we update $\theta_{\pi}$ with a mini-batch experiences $(\bm{s}^i,\bm{v}^i,c^i,\bm{s}^{i\prime}), i=1,...,N,$ randomly sampled from $R_{ac}$, 
\begin{equation}\label{eqn:anp_update}
\begin{array}{lrrl}
\theta_{\pi} = \theta_{\pi} -  \frac{\alpha_{\pi}}{N} \sum\limits_{i=1}^{N} \bigtriangledown_{\bm{v}} Q(\bm{s}^i,\bm{v}|\theta_Q)|_{\bm{v}=\pi(\bm{s}^i|\theta_{\pi})} \cdot \bigtriangledown_{\theta_{\pi}} \pi(\bm{s}^i), 
\end{array}
\end{equation}
where $\alpha_{\pi}$ is the learning rate of the Actor network. 
Notice that in the Actor network training, we only need the states $\bm{s}^i$ in the experiences and the actions $\bm{v}$ are generated by the Actor network with these states. 

%

\subsection{The Complete Algorithm}

Summarizing all the components, we present the pseudo-code in Algorithm \ref{alg:ACDQL}. 
The algorithm can be divided into two parts, the online action generating part (lines 3-7 and 18-19) and the asynchronous training part (lines 9-16). 
We briefly describe the steps as follows. 

At the beginning of each time slot $t$, the agent first uses the historic AR data $\bar{\bm{\lambda}}^t$ to predict the AR for the current time slot $\hat{\bm{\lambda}}^t$. 
Then, the system state is given by $\bm{s}^t=(\hat{\bm{\lambda}}^t,\bm{v}^{t-1})$ (line 4). 
The state is fed into the policy DNN to generate an analogue action $\tilde{\bm{v}}$, which after adding noise and rounding becomes the proto-action $\hat{\bm{v}}^t$ (line 5). 
After that, the agent generates the neighborhood set of $\hat{\bm{v}}$, $\mathcal{N}(\hat{\bm{v}})$ and refines the proto-action to obtain $\bm{v}^t$ (line 6). 
Then, the agent executes the action, i.e. sets the SBS modes (line 7). 

During the time span of time slot $t$, the agent updates the parameters of the networks (lines 9-16). 
The training of the APR DNN, CEN and AC networks can be done in parallel. 
We note that Critic network is only involved in the training. 
Notice that the training begins only when $t\geq N$, before which there are not enough experiences in the replay memories for one mini-batch. 

At the end of time slot $t$, the agent collects the true AR information $\bm{\lambda}^t$, measures the cost $c^t$ and observes the new state $\bm{s}^{\prime} = (\hat{\bm{\lambda}}^{t+1}, \bm{v}^t)$ (line 18). 
Meanwhile, it stores the experiences in the corresponding replay memories (line 19). 
If the memory is already full, the agent deletes the oldest experience to make room for the latest one.

\begin{algorithm}
\caption{Deep RL Based BS Activation Algorithm.}\label{alg:ACDQL}
\begin{algorithmic}[1]
\Require{Randomly initialize the network parameters $\theta_A$, $\theta_{\tilde{C}}$, $\theta_{\pi}$, $\theta_Q$, and set $\theta_{\pi}^T = \theta_{\pi}$, $\theta_{Q}^T = \theta_{Q} $}. 
\Require{Initialize replay memories $R_{A},R_{\tilde{C}},R_{ac}$.}
\Require{Initialize $\alpha_A,\alpha_{\tilde{C}}, \alpha_{\pi}, ~\alpha_{Q}$, $\sigma_n$, $\epsilon$ and $\bm{v}^0=[1,...,1]$.}
\For{$t=1:T_{steps}$} 
\State /* \textit{At the beginning of time slot $t$. } 
\State Decay $\alpha_{\pi}, ~\alpha_{Q},\alpha_{A},\alpha_{\tilde{C}}$, $\sigma_n$ and $\epsilon$. 
\State Predict the AR $\hat{\bm{\lambda}}^t = A(\bar{\bm{\lambda}}^t|\theta_A )$, set $\bm{s}^t=(\hat{\bm{\lambda}}^t,\bm{v}^{t-1})$. 
\State Generate a proto-action $\hat{\bm{v}}^t = f_r\left( \pi(\bm{s}^t|\theta_{\pi}) + \mathbb{N}(\sigma_n) \right)$. 
\State Refine $\hat{\bm{v}}^t$ with Eqn. (\ref{eqn:action_refine1}) to obtain action $\bm{v}^t$. 
\State Execute action $\bm{v}^t$. 
\State /* \textit{Train the networks during the span of time slot $t$. } 
\State Sample a mini-batch $(\bar{\bm{\lambda}}^i,\bm{\lambda}^i)$ from $R_{A}$. 
\State Update APR DNN parameters with Eqn. (\ref{eqn:arpp_update}). 
\State Sample a mini-batch $(\bm{\lambda}^i, \bm{v}^{i-1}, \bm{v}^i, c^i)$ from $R_{\tilde{C}}$. 
\State Update CEN parameters with Eqn. (\ref{eqn:lmp_update}). 
\State Sample a mini-batch $\left(\bm{s}^i, \bm{v}^i, c^i, \bm{s}^{\prime} \right)$ from $R_{ac}$.
\State Update Critic network parameters with Eqn. (\ref{eqn:cnp_update}). 
\State Update Actor network parameters with Eqn. (\ref{eqn:anp_update}). 
\State Update the target networks: 
\begin{equation}\label{eqn:tar_update}
\begin{array}{lrrl}
\theta_{Q}^T \leftarrow \tau \theta_Q + (1-\tau) \cdot \theta_{Q}^T, \\
\theta_{\pi}^T \leftarrow \tau \theta_{\pi} + (1-\tau) \cdot \theta_{\pi}^T.
\end{array}
\end{equation}
\State /* \textit{At the end of time slot $t$. } 
\State Observe actual AR $\bm{\lambda}^t$, actual cost $c^t $ and new state $\bm{s}^{\prime}$.
\State Store experience $(\bm{s}^t, \bm{v}^t, c^t, \bm{s}^{\prime})$ in $R_{ac}$, $(\bar{\bm{\lambda}}^t,\bm{\lambda}^t) $ in $ R_{A}$ and $ (\bm{\lambda}^t, \bm{v}^{t-1},\bm{v}^t, c^t ) $ in $R_{\tilde{C}}$. 
\EndFor
\end{algorithmic}
\end{algorithm}

\subsection{Implementation Issues}\label{subsec:implement}

The proposed algorithm operates in an online fashion and can adapt to time-varying traffic load patterns. 
To begin from a cold start, i.e. completely empty system, we first set the learning rates $\alpha_{\pi},\alpha_{Q},\alpha_{A}$ and $\alpha_{\tilde{C}}$ to relatively large values and then decrease them gradually to small fixed values. 
Notice that $\alpha_{\pi}$ should be larger than $\alpha_{Q}$, since the Actor network is trained with the gradient from the Critic network and can be viewed as in deeper layers. 
In contrast to conventional supervised training, in which the learning rate tends to 0 as training iteration grows, our online algorithm needs a non-zero learning rate to maintain adaptability to time-varying environment. 
To accelerate the learning process of the Actor-Critic networks and the training of the APR DNN and CEN, we carry out multiple training steps during a time slot. 
Notice that the replay memories are implemented with First-In-First-Out (FIFO) queues to keep only the freshest experiences. 
Trained with these fresh experiences, the proposed algorithm adapts to the time-varying environment in an online fashion.

We implement the algorithm using Python 3.6 and Google's machine learning library TensorFlow 1.0 \cite{tensorflow}, which provides high-level APIs for many machine learning procedures, e.g. gradient computing, activation function and computation graph building and visualization (tensorboard). 
All the DNNs are fully-connected networks. 
We adopt two additional enhancements to accelerate and stabilize the network training, namely, the Batch Normalization (BN) \cite{ioff_batc} and the Gradient Inverse (GI) \cite{GRAD_INV}. 
The DNN designs are summarized in Table \ref{table:net_conf}. 
We note that the network configurations are related to the system size of the HetNet, as will be shown in the simulation results. 
Nevertheless, we find that the performance of the algorithm is not very sensitive to the network configurations.

\begin{table}[h!]
\caption{Network Configurations}
\centering
\setlength\tabcolsep{3.5pt}
\begin{tabular}{c|c|c|c|c|c|c}
\hline
\multirow{2}{*}{ } & \multicolumn{2}{c|}{Layer 1} & \multicolumn{2}{c|}{Layer 2} & \multicolumn{2}{c}{Output layer} \\ \cline{2-7}
                   & Activation  & Size        & Activation & Size   & Activation 	& Size   \\ \hline
Actor              & BN+softplus & 200         & BN+relu    & 100    & tanh\footnotemark     & $B_s$ \\ \hline
Critic             & BN+softplus & 200         & BN+relu    & 100    & linear     	& 1      \\ \hline
APR DNN            & BN+tanh     & 200         & BN+tanh    & 100    & sigmoid    	& $B$    \\ \hline
CEN                & BN+tanh     & 200         & BN+tanh    & 100    & sigmoid    	& $1$  \\ \hline
\end{tabular}
\label{table:net_conf}
\end{table}
\footnotetext{We actually use $\frac{\text{tanh(x+2)}+1}{2}$.}

\section{Numerical Results}\label{sec:results}

In this section, we verify the superiority of DRAG and investigate how it is affected by the hyper parameters through numerical simulations. 
We first compare the performance of DRAG with the existing approaches and an upper bound benchmark. 
Then, we investigate the influence of the algorithmic parameters on the performances of the proposed algorithm.

\subsection{Simulation Settings}

Throughout this section, we consider a HetNet with one MBS in the network center and multiple (10 if not specified) SBSs within the coverage of the MBS. 
The coverage radii of the MBS and SBS are $1000$m and $100$m, respectively. 
The SBSs are distributed according to a Materns hard-core point process with a minimum distance of $200$m between each other. 
The maximum transmit powers of the MBS and SBS are 20W (43dBm) and 1W (30dBm), respectively. 
We summarize the other key system settings in the HetNet and the parameters of DRAG in Table \ref{table:system_set}. 
The power parameters are from real-network measurements \cite{deru_powe}. 

In our algorithm, we use Gaussian noise with standard deviation $\sigma_n$ in Eqn. (\ref{eqn:noi_exploration}). 
All the decaying parameters, i.e. $\sigma_n$, $\epsilon$, $\alpha_{\pi}, ~\alpha_{Q},~\alpha_A$ and $ ~\alpha_{\tilde{C}}$, decrease linearly from the upper limits to lower limits in 10000 time slots and keep fixed afterwards. 

\begin{table}[h!]
	\centering
	\setlength\tabcolsep{3.5pt}
	\caption{System settings and algorithmic parameters}
	\begin{tabular}{c|c|c|c}
    \hline
		$P_i^c,i\in \mathcal{B}_s$ & 160W  	& $P_i^l,i\in \mathcal{B}_s$  & 216W  \\ \hline
$P_m^l,m\in \mathcal{B}_m$  & 1080W 	& $B_s$      	& 10       	\\ \hline
		$\beta_d$  		& 50W/s 		& $\gamma$ 		& 0.9   	\\ \hline
		$\beta_s$  		& 100 Wh/time   & $\sigma_n$	& $0.5 - 0.05$	\\ \hline
		$h$          	&$4$ 			& $\epsilon$  	& $3 - 0.1$			\\ \hline
$|R_{A}|,|R_{\tilde{C}}|,|R_{ac}|$   	& 6000 		& $\alpha_{\pi}$  	& $5\cdot 10^{-3} - 8\cdot 10^{-4}$   	\\ \hline
		D & 1 			& $\alpha_{Q}$  	& $2\cdot 10^{-3} - 2\cdot 10^{-4}$   	\\ \hline
		$N$  		  	& 64   		& $\alpha_A$  	& $2\cdot 10^{-3} - 2\cdot 10^{-4}$  	\\ \hline
		$\tau$       	&$10^{-4}$	& $\alpha_{\tilde{C}}$  	& $2\cdot 10^{-3} - 2\cdot 10^{-4}$		\\ \hline		
	\end{tabular}
\label{table:system_set}
\end{table}


To measure the QoS degradation cost $c^t_d$, we adopt the metric of service delay, which is a closed-form function, from \cite{TACT} 
\begin{equation}\label{eqn:flow_delay}
\begin{array}{lrrl}
c_d^t(\bm{\rho}^t) = \sum\limits_{i\in \mathcal{B}} \frac{\rho^t_i}{1-\rho^t_i}, 
\end{array}
\end{equation}
where $\rho^t_i$ is the average system load measured at BS $i$ in time slot $t$.

In terms of benchmark algorithms, we consider the Q-Learning (QL), the classic Actor-Critic algorithm TACT \cite{TACT}, the optimization-based approach EECA \cite{zhua_ener} and the offline exhaustive search bound SOTA \cite{sonk_base}. 
The Q-Learning takes the AR of the last time slot $\bm{\lambda}^{t-1}$ as the state and uses the Boltzmann distribution for exploration, similar to TACT. 
Moreover, the optimization-based approach in \cite{zhua_ener} is referred to as EECA. 
For fair comparison, we add flow delay cost and switching cost to the objective function. 
The optimization problem in EECA is solved at the beginning of each time slot (half an hour) based on the traffic in the first minute of the time slot and is kept fixed until next time slot. 
On the other hand, SOTA exhaustively searches for the optimal action assuming that the load of the coming time slot is known noncausally and ignores the switching cost. 
This makes SOTA an unachievable lower bound in general, since the switching cost is non-zero in the long run. 
In QL and TACT, the continuous AR values are quantized into 5 intervals. 
Note that since the AR varies across time slots, we use the daily cost, which is the average of costs of all time slots of a day, to measure the performance. 
For unified comparison, the results are all normalized to the cost of the case when all SBSs are active.

\subsection{Performance vs AR Pattern}

\begin{figure}[h!]
\centering
\includegraphics[width=\figwidth]{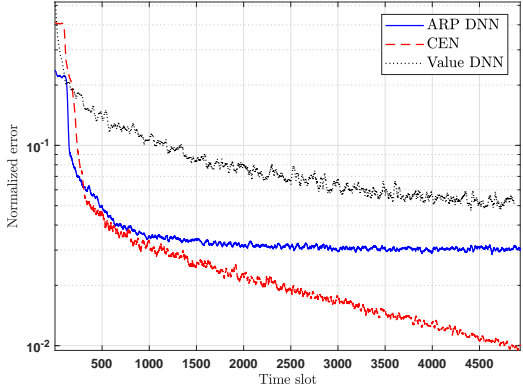}
\caption{Training errors of the component networks.} \label{fig:error}
\end{figure}

We first show the convergence performance (training speed) of the APR DNN, CEN and the Value DNN in Fig. \ref{fig:error}. 
The errors are the normalized training errors measured with the mini-batch, e.g. for the cost $c$, $e_c=\mathbb{E}\left[ \frac{||\hat{c}^i-c^i||_2}{||c^i||_2}\right]$. 
It can be observed that errors of the APR DNN and CEN quickly decrease to very low while that of the Value DNN decreases much slower. 
This implies that the former two networks converge much faster than the Value DNN. 
For example, at time slot 1000, the errors of the APR DNN and CEN are around $3\%$, while that of the Value DNN is around $10\%$. 
The reason is that the APR DNN and CEN are trained with accurate data samples (albeit with noise) in a supervised fashion, while the Value DNN is trained with inaccurate samples in an unsupervised way. 
Notice that the error of the APR DNN does not diminish to 0. 
This is due to the average noise in the AR data. 

\subsubsection{Stationary AR Pattern}\label{subsubsec:sar}

We first compare DRAG with the existing methods when the AR pattern is stationary. 
For the AR traces, we first capture a basic AR pattern by uniformly retrieving 48 sample points from the daily load pattern in \cite{peng_traf}. 
Each of the sample point denotes a time slot that lasts half an hour. 
The AR pattern for each SBS is a randomly scaled and shifted version of the basic AR pattern, where the scale factor and time slots shift are randomly selected from $[0.6,1]$ and $[-8,8]$, respectively. 
Randomness is then added to the ARs with a correlated Gaussian noise generated from an Ornstein$-$Uhlenbeck process with $\theta=0.05,\sigma=0.03$, where $\theta$ measures the independence between neighboring noise and $\sigma$ is the Gaussian standard deviation. 
Under this setting, the noise process resembles a random walk and usually reaches a maximum deviation of around $\pm 15\%$. 
All the results are obtained by averaging 20 AR traces, each of which lasts $2\cdot 10^4$ time slots (416 days).

\begin{figure}[h!]
\vspace{-0.2cm}
\centering
\includegraphics[width=\figwidth]{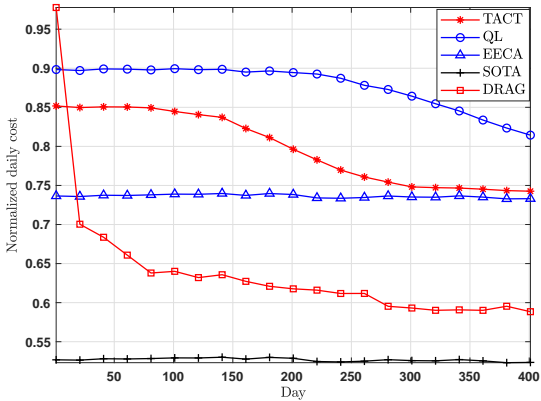}
\caption{Normalized daily cost curves with stationary AR pattern.} \label{fig:comparison_sp}
\end{figure}

The result is shown in Fig. \ref{fig:comparison_sp}, in which the data sequences have been smoothened by moving average with window size of 10 days. 
We can see that in the long run, the proposed algorithm DRAG outperforms the other learning methods significantly. 
Moreover, it also surpasses the optimization based approach EECA. 
Despite the random initialization, DRAG learns very fast and quickly approaches the lower bound, SOTA.

The main reasons for the poor performance of the tabular-based learning methods, i.e. TACT and QL, are as follows. 
Firstly, they have a high-dimensional state-action space and have low exploration efficiency in an environment with varying AR. 
In fact, in TACT, there are $2^{10}$ actions and $5^{10}\cdot 2^{10}$ state-action pairs, while only 0.0004$\%$ of the state-action pairs and 0.08$\%$ of the states have been visited in 416 days (20000 time slots). 
We find that they perform much better in fixed AR scenarios, as shown in the original paper \cite{TACT}. 
Secondly, they have poor generalization ability among similar state-action pairs, e.g. non-visited states simply have zero values. 
The low exploration and poor generalization together render the learning process of these tabular-based methods very slow. 
These problems become even worse when the system grows larger, as will be shown later. 
As for EECA, its performance gap with SOTA comes from the uncertainty of the traffic and its inability to foresee the traffic. 
Since it only uses the traffic information of the first minute of each time slot, it may underestimate the traffic when the traffic is in the increasing slop or overestimate when the traffic is in the decreasing trend.

We note that the relatively good performance of EECA comes at a cost of high computational time. 
In fact, EECA has a computational complexity of $O(nB\cdot 2^{B})$, where $n$ is the number of user groups and $B$ is the number of BSs, including MBSs and SBSs. 
Although the refined algorithm in \cite{zhua_ener} significantly reduces the complexity above, it still takes around 100 seconds in a scenario with 2 MBS and 10 SBSs and would take much longer time in larger systems. 
On the other hand, to generate a policy in exploration, the TBRL methods usually need $O(2^{B_s})$ computations, which is not scalable in large systems. 
In contrast, DRAG only needs a forward evaluation of the DNN, which needs only some matrix multiplications, to generate a policy and $O(B_s)$ computations to refine the action. 
Note that the DNN training has the same complexity as forward evaluation and can be carried out amid the time slots. 
As will be shown later, a sub-linear growth in the DNN size is sufficient to provide enough capacity scaling. 
Therefore, the complexity of DRAG is at most $O(B_s^2)$.

\subsubsection{Slowly Varying AR Pattern}

We also investigate the ability of DRAG to adapt to varying AR patterns. 
In this case, the settings are similar to that of the stationary AR case except that after every 100 days, the ARs of the SBSs are re-scaled and re-shifted.

\begin{figure}[h!]
\vspace{-0.2cm}
\centering
\includegraphics[width=\figwidth]{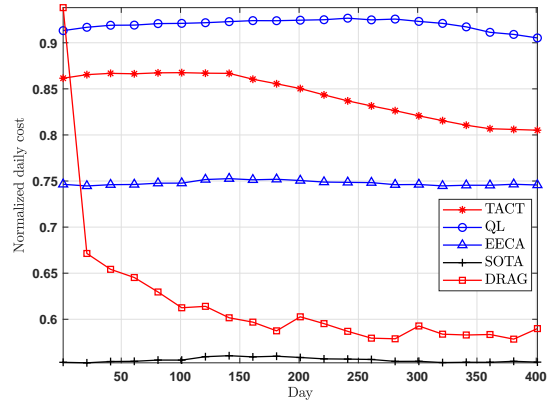}
\caption{Normalized daily cost curves with varying AR pattern.} \label{fig:comparison_fvp_2}
\end{figure}

We can see from Fig. \ref{fig:comparison_fvp_2} that again, DRAG significantly outperforms TACT and Q-learning methods. 
At each AR pattern transition (100, 200, 300 and 400 days), the performance of DRAG deteriorates due to the sudden changes. 
However, it can keep the deterioration low and regain the high performance very quickly. 
In contrast, the deterioration in the final performance of QL and TACT is much greater, since the sudden change in AR brings new states (AR combinations), which further sparsifies the learning experience.

\subsubsection{Noise Intensity}

\begin{figure}[h!]
\vspace{-0.2cm}
\centering
\includegraphics[width=\figwidth]{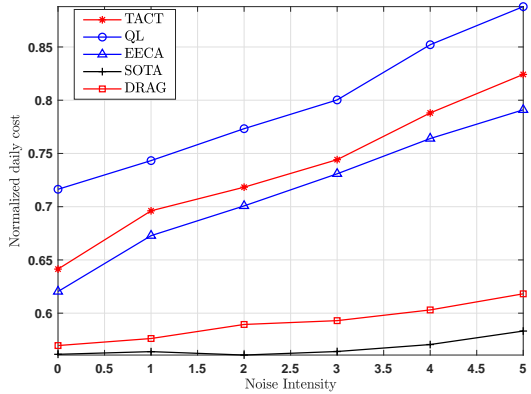}
\caption{Normalized daily cost curves with different noise intensities.} \label{fig:noi_int_per_comp}
\end{figure}

We further demonstrate the learning capability of DRAG by varying the noise intensity. 
In particular, we intent to show that the stronger the traffic pattern is, the faster and more effectively that DRAG can learn. 
Nevertheless, DRAG also possesses a good resistance to uncertainty. 
To show that, we present the simulation results with different noise intensities in the traffic trace. 
In the Ornstein$-$Uhlenbeck process, we vary the standard deviation $\sigma$ of the Gaussian noise from 0 to 0.05. 
The corresponding average amplitudes of the noise curve varies from 0 to $0.3$. 
The results are shown in Fig. \ref{fig:noi_int_per_comp}, in which the normalized daily costs are the mean of the final 20 days and averaged from 20 instances.

It can be seen that DRAG is robust to the noise intensity while the other three algorithms are much more sensitive. 
The reason for performance degradation of QL and TACT is that they use the AR of the previous time slot as state. 
Thus, when the noise becomes stronger, the learned experience in the action space becomes sparser and the reinforcement effect (learned pattern) is weakened. 
On the other hand, the sensitivity of EECA to the noise intensity is due to the fact that it optimizes over the AR of the first traffic realization in each time slot. 
When the noise intensity increases, the correlation between traffic intensity in time decreases. 
In contrast, with the help of action refinement the DNN's generalization ability, DRAG is much more robust to the noise intensity.

\subsection{System Scale and Hyper Parameters}

\subsubsection{System Scale}

We investigate the performance with respect to the system scale. 
Specifically, we vary the number of SBSs from 6 to 16. 
For all approaches, the system is run by 400 days with the stationary AR traces specified in \ref{subsubsec:sar}. 
The results are obtained by averaging the daily cost of the final 20 days. 

\begin{figure}[h!]
\vspace{-0.2cm}
\centering
\includegraphics[width=\figwidth]{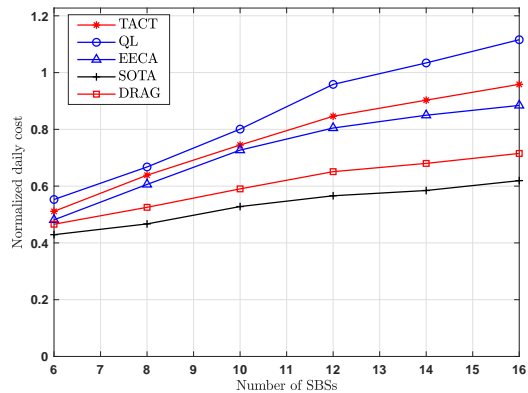}
\caption{Performance versus number of SBSs.} \label{fig:prob_scal_per_comp}
\end{figure}

In Fig. \ref{fig:prob_scal_per_comp}, the performance of tabular learning approaches degradates significantly when the system size grows. 
This is because when the number of SBSs increases, the learned experiences become sparser due to the larger action and state-action spaces. 
In contrast, our proposed algorithm shows consistent performance with respect to the system scale, i.e. the performance gap between DRAG and SOTA is consistently small.

\subsubsection{Effects of Action Refinement}

\begin{figure}[h!]
\vspace{-0.2cm}
\centering
\includegraphics[width=\figwidth]{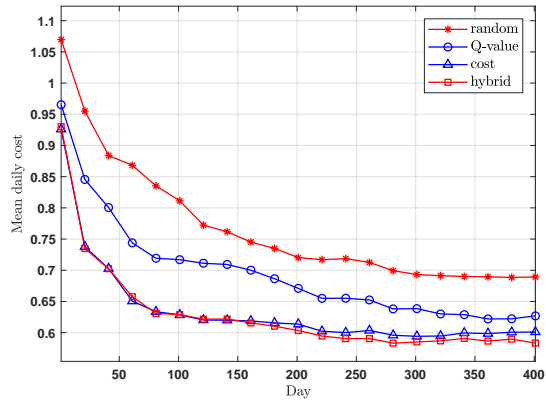}
\caption{Performance of different exploration methods.} \label{fig:action_refinement}
\end{figure}

To verify the performance of the proposed action refinement procedure, we compare the daily costs with pure random exploration (default in DDPG), Q-value based action refinement (the Wolpertinger policy \cite{action_refine}), the proposed cost-greedy and the hybrid method ($\epsilon$-cost-greedy), in Fig. \ref{fig:action_refinement}. 
It can be seen that the cost-greedy and the hybrid exploration procedure significantly accelerate the learning process and outperform the purely random and Q-value based explorations. 
The reason is that initially, the estimation of the Q-value is very inaccurate until the 300th day, while the cost estimation quickly becomes accurate in 50 days (see Fig. \ref{fig:error}). 
Compared with the Q-value estimation, the accurate cost estimation provides better guidance for exploration and action evolution. 
The closeness of the performance of cost-greedy and the hybrid in the initial stage (before day 150) is due to that the hybrid essentially uses the cost-greedy in this period, i.e. $\epsilon(t)\geq 1$. 

\subsubsection{Complexity and Scalability}

\begin{figure}[h!]
\vspace{-0.2cm}
\centering
\includegraphics[width=\figwidth]{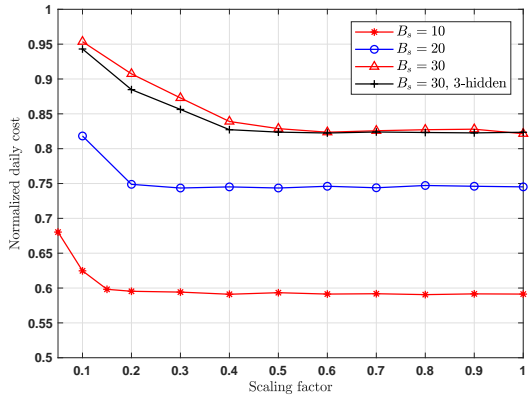}
\caption{Performance versus DNN size.} \label{fig:net_size_scale}
\end{figure}

In the above, we have shown that DRAG performs consistently well in different system settings. 
Here, we investigate what it takes to achieve the good performance when the system size changes, or equivalently, how the Required Network Size (RNS) scales with the system size. 
To this end, we vary the network size and investigate its impact on the daily cost for three cases $B_s=10, ~B_s=20$ and $B_s=30$, respectively. 
We refer to the numbers of neural units in the layers as network width and the number of hidden layers as network depth. 
In the following, the network size is varied in both the width and depth dimension based on a basic setting specified in \ref{subsec:implement}, i.e. all networks have hidden layer width as $[200, 100]$. 
For the basic setting of the three hidden layers case, we replicate the settings (number of neurons and activation functions) of the second layer to the third layer. 
We scale the network width by a factor $k$, i.e. $k\cdot [200, 100]$, and compare the achieved daily cost. 
For each of the network size setting, we train the agent for $2\cdot 10^4$ time slots with the stationary AR traces. 
The daily cost is obtained by averaging the final 20 days' cost of 20 instances. 
In Fig. \ref{fig:net_size_scale}, the normalized reward is obtained by normalizing the mean daily reward to that of SOTA, which is an upper bound.

From Fig. \ref{fig:net_size_scale}, we see that the achieved daily cost quickly converges when the scaling factor increases. 
The convergence implies that the network size is large enough for the agent to learn the best policy. 
Specifically, for case $B_s=10$, a scale of 0.15, or equivalently $[30, 15]$, is sufficient, while for $B_s=20$ and $B_s=30$ (two-layer), the convergent scales are 0.3 ($[60, 30]$) and 0.5 ($[100, 50]$), respectively. 
It is worth noting that when we increase the network depth (use three hidden layers), the convergent scale factor for $B_s=30$ decreases to 0.4 ($[80, 40, 40]$). 
This means that the increasing in depth also increases the DNN's capacity and alleviates the demand in network width. 

It is seen in Fig. \ref{fig:net_size_scale} that the RNS does not increase fast with the system size $B_s$. 
In fact, Fig. \ref{fig:net_size_scale} implies that the RNS increases almost linearly with $B_s$, i.e. $RNS=O(B_s)$. 
Moreover, when combined with increased depth, we can further decrease the order of the RNS. 
The above observation implies that the complexity of the proposed deep learning algorithm is $O(B_s^2)$, which hopefully can be further reduced to $O(B_s)$ with the increase of network depth. 
In contrast, the tabular-based learning methods have complexity of $O(2^{B_s})$ (in the policy generation step during exploration).

\section{Conclusions}\label{sec:conslusion}

In this paper, we have developed a deep reinforcement learning-based SBS activation approach for energy saving in large HetNets. 
We first formulate the SBS activation problem as an MDP to minimize the cumulative energy consumption, with the switching cost and service delay considered. 
To solve the MDP, we adopt the Actor-Critic reinforcement learning framework and propose to use DNNs to approximate the policy and value functions in the framework. 
Meanwhile, we use a DNN to explicitly predict the arrival rates, which exploits the temporal and spatial correlations among SBS traffics. 
Furthermore, we utilize a CEN to estimate the cost given state and action. 
With this, we have proposed a novel action refinement procedure to aid exploration and accelerate learning. 
As such, the algorithm is completely model-free, data-driven and can adapt to varying network environment. 
By extensive simulations, we demonstrate that the proposed DRAG algorithm significantly outperforms the existing learning methods in both stationary traffic case and varying traffic case. 
Moreover, we show that the algorithm can scalably extend to large systems, as opposed to the existing tabular-learning based methods.

{}


\begin{thebibliography}{plain}

\bibitem{glob_anin}
Global Action Plan, An inefficient truth, \url{http://www.globalactionplan.org.uk/}, Global Action Plan Rep., 2007.



\bibitem{TACT}
R. Li, Z. Zhao, X. Chen, J. Palicot, ``TACT: A transfer actor-critic learning framework for energy saving in cellular radio access networks," \textit{IEEE Trans. on Wireless Commun.}, vol. 13, no. 4, pp. 2000-2011, Apr. 2014.


\bibitem{peng_traf}
C. Peng, S. B. Lee, S. Lu, et al., ``Traffic-driven power saving in operational 3G cellular networks," in \textit{ACM Proceedings of the 17th Annual International Conference on Mobile Computing and Networking}, Las Vegas,  USA, Sep. 2011.





\bibitem{mars_opti}
M. A. Marsan, L. Chiaraviglio, D. Ciullo, et al., ``Optimal energy savings in cellular access networks," in \textit{ICC Workshops}, Dresden, Germany, Aug. 2009.


\bibitem{kong_rein1}
P. Y. Kong and D. Panaitopol, ``Reinforcement learning approach to dynamic activation of base station resources in wireless networks," in \textit{IEEE 24th International Symposium on Personal Indoor and Mobile Radio Communications}, London, UK, Nov. 2013.


\bibitem{suar_anov}
L. Suarez, L. Nuaymi, and J. M. Bonnin, ``An overview and classification of research approaches in green wireless networks," \textit{Eurasip Journal on Wireless Communications and Networking}, vol. 2012, no. 1, Dec. 2012.


\bibitem{alca_9900}
Alcatel-Lucent, “9900 wireless network guardian,” White paper, Aug. 2012. [Online]. Available: \url{http://www3.alcatel-lucent.com/wps/DocumentStreamerServlet?LMSG_CABINET=Docs_and_Resource_Ctr&LMSG_CONTENT_FILE=White_PapersGuardian_EN_Tech_WhitePaper.pdf}


\bibitem{moro_impr}
S. Morosi, P. Piunti, E. Del Re, ``Improving cellular network energy efficiency by joint management of sleep mode and transmission power," in \textit{Tyrrhenian International Workshop on Digital Communications-Green ICT}, Genoa, Italy, Sep. 2013.




\bibitem{jiac_reso}
C. Jia and T. J. Lim, ``Resource partitioning and user association with sleep-mode base stations in heterogeneous cellular networks," \textit{IEEE Transactions on Wireless Communications}, vol. 14, no. 7, pp. 3780-3793, Jul. 2015.

\bibitem{sung_ener}
S. Cho and W. Choi, ``Energy-efficient repulsive cell activation for heterogeneous cellular networks," \textit{IEEE Journal on Selected Areas in Communications}, vol. 31, no. 5, pp. 870-882, May 2013.


\bibitem{wuji_traf}
J. Wu, S. Zhou, and Z. Niu, ``Traffic-aware base station sleeping control and power matching for energy-delay tradeoffs in green cellular networks," \textit{IEEE Trans. on Wireless Commun.}, vol. 12, no. 8, pp. 4196-4209, Aug. 2013.


\bibitem{wuji_base}
J. Wu, Y. Bao, G. Miao, et al., ``Base-station sleeping control and power matching for energy–delay tradeoffs with bursty traffic," \textit{IEEE Trans. on Vehicular Technology}, vol. 65, no. 5, pp. 3657-3675, May 2016.




\bibitem{sohy_ener}
Y. S. Soh, T. Q. S. Quek, M. Kountouris, et al., ``Energy efficient heterogeneous cellular networks," \textit{IEEE J. Sel. Areas Commun.}, vol. 31, no. 5, pp. 840-850, Apr. 2013.


\bibitem{wild_cogn}
M. Wildemeersch, T. Q. S. Quek, C. H. Slump, et al., ``Cognitive small cell networks: Energy efficiency and trade-offs," \textit{IEEE Trans. on Commun.}, vol. 61, no. 9, pp. 4016-4029, Jul. 2013.

\bibitem{taba_down}
H. Tabassum, U. Siddique, E. Hossain, et al., ``Downlink performance of cellular systems with base station sleeping, user association, and scheduling," \textit{IEEE Trans. on Commun.}, vol. 13, no. 10, pp. 5752-5767, Jul. 2014.







\bibitem{liao_base}
W. C. Liao, M. Hong, and Y. F. Liu, ``Base station activation and linear transceiver design for optimal resource management in heterogeneous networks," \textit{IEEE Transactions on Signal Processing}, vol. 62, no. 15, pp. 3939-3952, Aug. 2014.

\bibitem{zhua_ener}
B. Zhuang, D. Guo, and Michael L. Honig, ``Energy-efficient cell activation, user association, and spectrum allocation in heterogeneous networks," \textit{IEEE Journal on Selected Areas in Communications}, vol. 34, no. 4, pp. 823-831, Apr. 2016.


\bibitem{sonk_base}
K. Son, H. Kim, Y. Yi, ``Base station operation and user association mechanisms for energy-delay tradeoffs in green cellular networks," \textit{IEEE J. Sel. Areas Commun}, vol. 29, no. 8, pp. 1525-1536, Sep. 2011.








\bibitem{cais_acro}
S. Cai, L. Xiao, H. Yang, et al., ``A cross-layer optimization of the joint macro-and picocell deployment with sleep mode for green communications," in \textit{IEEE Wireless and Optical Communication Conference (WOCC)}, Chongqing, China, Dec. 2013.



\bibitem{sake_opti}
L. Saker, S. E. Elayoubi, R. Combes, et al., ``Optimal control of wake up mechanisms of femtocells in heterogeneous networks," \textit{IEEE J. Sel. Areas Commun.}, vol. 30, no. 3, pp. 664-672, Mar. 2012.


\bibitem{DDPG}
T. P. Lillicrap, J. J. Hunt, A. Pritzel, et al., ``Continuous control with deep reinforcement learning," in \textit{International Conference on Learning Representations}, San Juan, Puerto Rico, May 2016.




\bibitem{deru_powe}
M. Deruyck, W. Joseph, and L. Martens, ``Power consumption model for macrocell and microcell base stations," \textit{Transactions on Emerging Telecommunications Technologies}, vol. 25, no. 3, pp. 320-333, 2014.


\bibitem{ashr_slee}
I. Ashraf, F. Boccardi, and L. Ho, ``Sleep mode techniques for small cell deployments," \textit{IEEE Communications Magazine}, vol. 49, no. 8, Aug. 2011.












\bibitem{sutt_rein}
R. S. Sutton and A. G. Barto, Reinforcement learning: An introduction. Vol. 1. No. 1. Cambridge: MIT press, 1998.



\bibitem{igro_effi}
I. Grondman, M. Vaandrager, L. Busoniu, et al., ``Efficient model learning methods for actor–critic control,” \textit{IEEE Transactions on Systems, Man, and Cybernetics, Part B (Cybernetics)}, vol. 42, no. 3, pp. 591–602, Jun. 2012.


\bibitem{jpet_natu}
J. Peters and S. Schaal, ``Natural actor–critic,” \textit{Neurocomputing}, vol. 71, nos. 7–9, pp. 1180–1190, Mar. 2008.




\bibitem{q_learn}
Christopher John Cornish Hellaby Watkins, \textit{Learning from delayed rewards}, Diss. King's College, Cambridge, 1989.


\bibitem{SARSA}
G. A. Rummery and M. Niranjan, On-line Q-learning using connectionist systems. Vol. 37. University of Cambridge, Department of Engineering, 1994. 

\bibitem{AC}
A. G. Barto, R. S. Sutton, and C. W. Anderson, ``Neuronlike adaptive elements that can solve difficult learning control problems," \textit{IEEE Transactions on Systems, Man, and Cybernetics}, vol. SMC-13, no. 5, pp. 834-846,  Sep.-Oct. 1983.


\bibitem{weiy_user}
Y. Wei, F. R. Yu, M. Song, Z. Han, ``User Scheduling and Resource Allocation in HetNets With Hybrid Energy Supply: An Actor-Critic Reinforcement Learning Approach," \textit{IEEE Trans. on Wireless Commun.}, vol. 17, no. 1, pp. 680-692, Jan. 2018.


\bibitem{DDQN}
H. Van Hasselt, A. Guez, and D. Silver, ``Deep Reinforcement Learning with Double Q-Learning," in \textit{Proceedings of the Thirtieth AAAI Conference on Artificial Intelligence}, Phoenix, USA, Feb. 2016.




\bibitem{DQN}
V. Mnih, K. Kavukcuoglu, D. Silver, et al., ``Human-level control through deep reinforcement learning," \textit{Nature}, vol. 518.7540, Feb. 2015.


\bibitem{maoh_neur}
H. Mao, R. Netravali, M. Alizadeh, ``Neural adaptive video streaming with pensieve," in \textit{Proceedings of the Conference of the ACM Special Interest Group on Data Communication}, Los Angeles, USA, Aug. 2017.



\bibitem{horn_mult}
K. Hornik, M. Stinchcombe, H. White, ``Multilayer feedforward networks are universal approximators," \textit{Neural networks}, vol. 2, no. 5, pp. 359-366, 1989.



\bibitem{plap_para}
M. Plappert, R. Houthooft, P. Dhariwal, et al., ``Parameter space noise for exploration," in \textit{International Conference on Learning Representations}, Vancouver, Canada, Apr. 2018.


\bibitem{action_refine}
G. Dulac-Arnold, R. Evans, H. van Hasselt, et al., ``Deep reinforcement learning in large discrete action spaces," in \textit{International Conference on Machine Learning}, New York, USA, Jun. 2016.




\bibitem{DPG}
D. Silver, G. Lever, N. Heess, et al., ``Deterministic policy gradient algorithms," in \textit{International Conference on Machine Learning}, Beijing, China, Jun. 2014.


\bibitem{ioff_batc}
S. Ioffe and C. Szegedy, ``Batch normalization: Accelerating deep network training by reducing internal covariate shift," in \textit{International Conference on Machine Learning}, Lille, France, Jul. 2015.


\bibitem{GRAD_INV}
M. Hausknecht and P. Stone, ``Deep reinforcement learning in parameterized action space," in \text{International Conference on Learning Representations}, San Juan, Puerto Rico, May 2016.












\bibitem{tensorflow}
“Tensorflow.org.” \url{https://www.tensorflow.org/}.




\end{thebibliography}
\end{document}